# Coherently Assisted Wireless Power Transfer Through Barely Transparent Barriers

Alex Krasnok

*Photonics Initiative, Advanced Science Research Center, City University of New York, NY 10031, USA*

To whom correspondence should be addressed: akrasnok@gc.cuny.edu

## Abstract

Wireless power transfer (WPT) technologies are becoming ubiquitous, but unlike wire technologies, they do not allow efficient energy transfer through barely transparent barriers. The proposed solutions to this issue rely on resonantly enhanced transparency of specially designed obstacles and hence are narrowband and require using artificially tailored materials. In this work, we suggest a fundamentally different approach to wireless power transfer through such barriers based on tailoring of a coherent auxiliary wave with a certain amplitude and phase from the side where the energy should be delivered to. This wave facilitates signal transmission and reflects back to the receiver allowing archiving of 100% energy transfer efficiency even for barely transparent barriers. In contrast to the traditional solution, this approach is general, applicable to lossless and lossy barriers and does not require cost-ineffective and complicated artificial structures.

## Introduction

Wireless power transfer (WPT) proposed at the beginning of the 20th century by N. Tesla [1] relied predominantly on far-field radiation and hence required high intensities, stability concerns, massive highly directive antennas at both transmitting and receiving ends [2]. Despite that this technology eventually led to wireless information transfer, the power transfer was developing in a different way with the use of wires. The WPT technologies had experienced a rebirth in 2007 when the group of Marin Soljačić has invented a principally new and more efficient approach based on the near-field coupling of two magnetic resonators [3,4]. The demonstration that the WPT efficiency between two metallic coils placed at the distance over 2 m can overcome 45% in the kHz range in this work [3] has given rise to tremendous research efforts in this promising direction [2,5–10] and resulted in the implementation of WPT systems for many vital applications, including consumer electronics, implanted devices [11], electric vehicles [12], to name just a few [6].

In applications, it is often needed to transfer energy through barely transparent barriers. In radio and microwave ranges this is associated with a problem of wireless transfer through, for example, walls of reinforced concrete buildings, Figure 1(a). Also, it is often needed to make walls or barriers transparent only at specific frequencies or circumstances for the sake of electromagnetic protection or security. The proposed solutions to this issue are based on resonantly enhanced transparency of specially designed barriers, e.g., perforated subwavelength holes or slits [13–15]. However, this approach operates in a narrow spectral range that limits the number of possible WPT channels and requires specifically tailored artificial materials making this technology cost-ineffective.

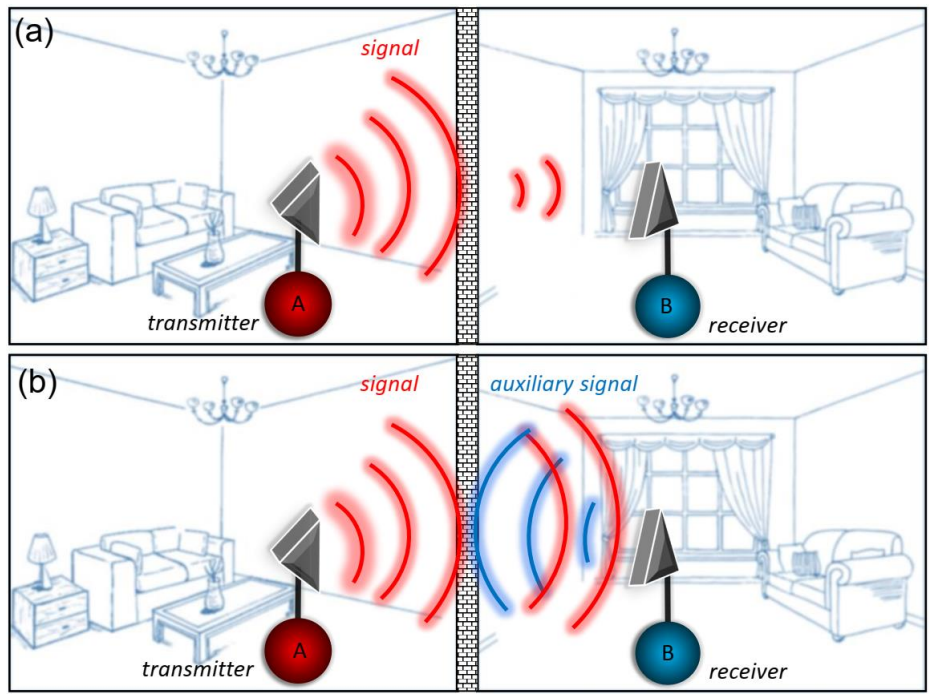

**Figure 1**. (a) Illustration of a problem of wireless power transfer through barely transparent barriers on the example of walls of reinforced concrete buildings. The low transmission through the wall would require either specific designing of the wall structure (e.g., periodically arranged subwavelength holes) or increase of the radiation energy. (b) The essence of the proposed approach in this work: the wall is irradiated by an auxiliary wave coherent to the initial signal. In this case,

the task of energy transfer lies entirely on the design of this auxiliary signal and can be fulfilled for a barrier of any type. The auxiliary wave is perfectly reflected back and hence not lost. In lossless case, the WPT efficiency can reach 100% for any barely transparent barrier.

In this work, we suggest a fundamentally novel approach to efficient wide-band wireless power transfer through barely transparent barriers. In contrast to the traditional solution, this approach is general and does not require using artificial materials and can be applied to a barrier of any type and in any spectral range. The approach is based on coherent excitation of a barrier by a specially designed auxiliary wave with a certain amplitude and phase from the side where the energy should be delivered to, Figure 1(b). This auxiliary wave not only helps the initial signal travel through the barrier but also gets fully reflected back ensuring 100% transfer efficiency in a lossless case. Firstly, we analyze the issue of energy transfer through barriers analytically in terms of scattering (S) matrix and derive the general equation, which allows treating the barrier of any type in the same way. Then, we demonstrate the applicability of the proposed approach for two different barriers: a Fabry-Perot high-index dielectric resonator and more complicated nonresonant structure in the form of periodically arranged subwavelength holes (metasurface). Finally, we demonstrate the usefulness of the proposed coherent approach in the case of the presence of dissipative losses.

This work is inspired by recent studies in coherent electromagnetics and photonics, which have been triggered by the implementation of coherent perfect absorption (CPA) [16–19] and subsequent realization of ultimate all-optical light manipulation [20], and enhanced wireless power transfer [21,22].

## Results and Discussion

The general problem of electromagnetic wave transmission through a linear system can be formulated in terms of the S-matrix approach in the following way. The total field in a system can be represented by the superposition of incoming ($E_m^+$) and outgoing ($E_m^-$) waves (e.g., plane waves or spherical harmonics) in different scattering channels, $E_i = \sum_m s_m^+ E_m^+$, $E_o = \sum_m s_m^- E_m^-$, with the amplitudes $s^+ = \{s_1^+, s_2^+, ...\}$ and $s^- = \{s_1^-, s_2^-, ...\}$. The amplitudes are usually normalized such that $|s_m^+|^2$ and $|s_m^-|^2$ correspond to the energy of the incoming and outgoing waves in a particular channel $m$. Thus, the formal solution to the general scattering problem is to find the output scattering amplitudes, which are linked to the input ones through the S-matrix ($\hat{S}$), $s^- = \hat{S}s^+$. For example, in a single port system, the $\hat{S}$ matrix coincides with the reflection coefficient ($r$). In a two-port reciprocal system, the scattering matrix is $\hat{S} = \begin{pmatrix} r_{11} & t \\ t & r_{22} \end{pmatrix}$, where $r_{ii}$ and $t$ stand for corresponding reflection and transmission coefficients, Figure 2(a).

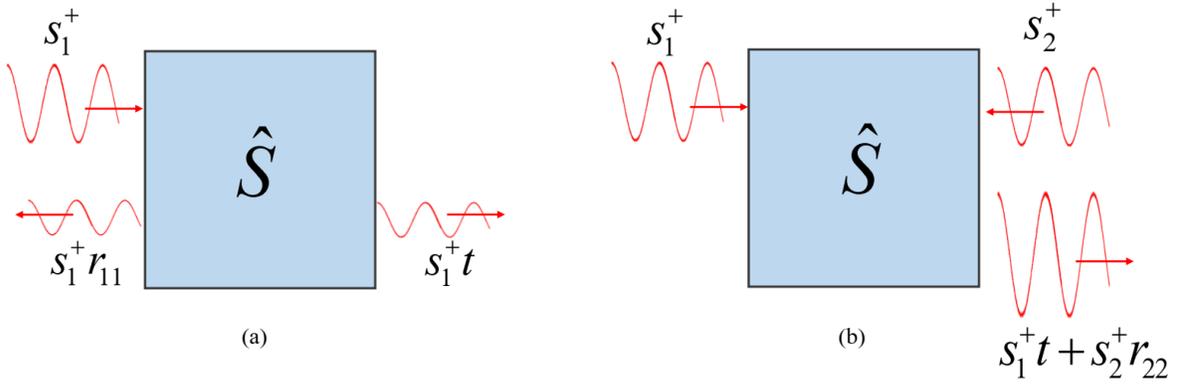

(a) (b)

**Figure 2**. (a) One-port excitation of a two-port system. (b) Two-port excitation leads to the cancelation of the reflected wave to port #1 and full transmission to port #2.

Knowledge of the S-matrix of a general two-port system excited from both ports allows finding the net energy flow to the port #2, $|s_1^+ t + s_2^+ r_{22}|^2$, Figure 2(b). This energy flow is only a part of total energy, which goes to the system: $|s_1^+|^2 + |s_2^+|^2$. Thus, the part of the total energy going to port #2 is defined by the normalized transferred power ($\Sigma(s_2^+) \geq 0$)

$$\Sigma(s_2^+) = \frac{|s_1^+ t + s_2^+ r_{22}|^2}{\left(|s_1^+|^2 + |s_2^+|^2\right)}. \tag{1}$$

Next, we assume that the signal wave (port #1) has the unit amplitude and oscillate at the frequency $\omega$, i.e., $s_1^+ = 1e^{i\omega t}$. According to our strategy, we suppose the second auxiliary wave at port #2 oscillates at the same frequency but has the amplitude $a$ and relative phase $\varphi$, $s_2^+ = ae^{i\omega t + i\varphi}$. For these waves, we can find the general equation for the normalized transferred power:

$$\Sigma(a, \varphi) = \frac{|t|^2 + a^2 |r_{22}|^2 + 2a \operatorname{Re}\left[ tr_{22}^* e^{-i\varphi} \right]}{(1 + a^2)}. \tag{2}$$

This formula is fair for a wide class of reciprocal two-port systems and can be easily generalized to nonreciprocal systems. Below, we are also interested in the power transfer enhancement relative to the case without auxiliary wave $[\Sigma(s_2^+ = 0) = |t|^2]$, $B(s_2^+) = \Sigma(s_2^+)/|t|^2$.

To begin with, let us consider a simple Fabry-Perot cavity in the form of a dielectric slab with relative dielectric permittivity $\varepsilon = 100$ situated in the air, Figure 3. Since this layout is general for many spectral ranges including GHz, THz, microwaves and optics, below we use the dimensionless units. The only one dimension parameter of this one-dimensional problem is the thickness, $L$, which in our calculations has been taken to be 405 nm. Using traditional T-matrix approach [23], we analytically calculate reflection (R) and transmission (T) coefficients of this slab, Figure 3(a, solid curves). For comparison, the results of numerical simulation in CST Microwave Studio are also presented by red and blue dots. We see that between two transmission Fabry-Perot resonances with vanishing $R$, there is a stop zone between $k_0 L = 0.65$ and $k_0 L = 0.7$ ($R \approx 0.95$ at $k_0 L = 0.675$), where $k_0 = \omega/c$, wavenumber in the free space. Consequently, the transmission coefficient in this area is very weak (~5%).

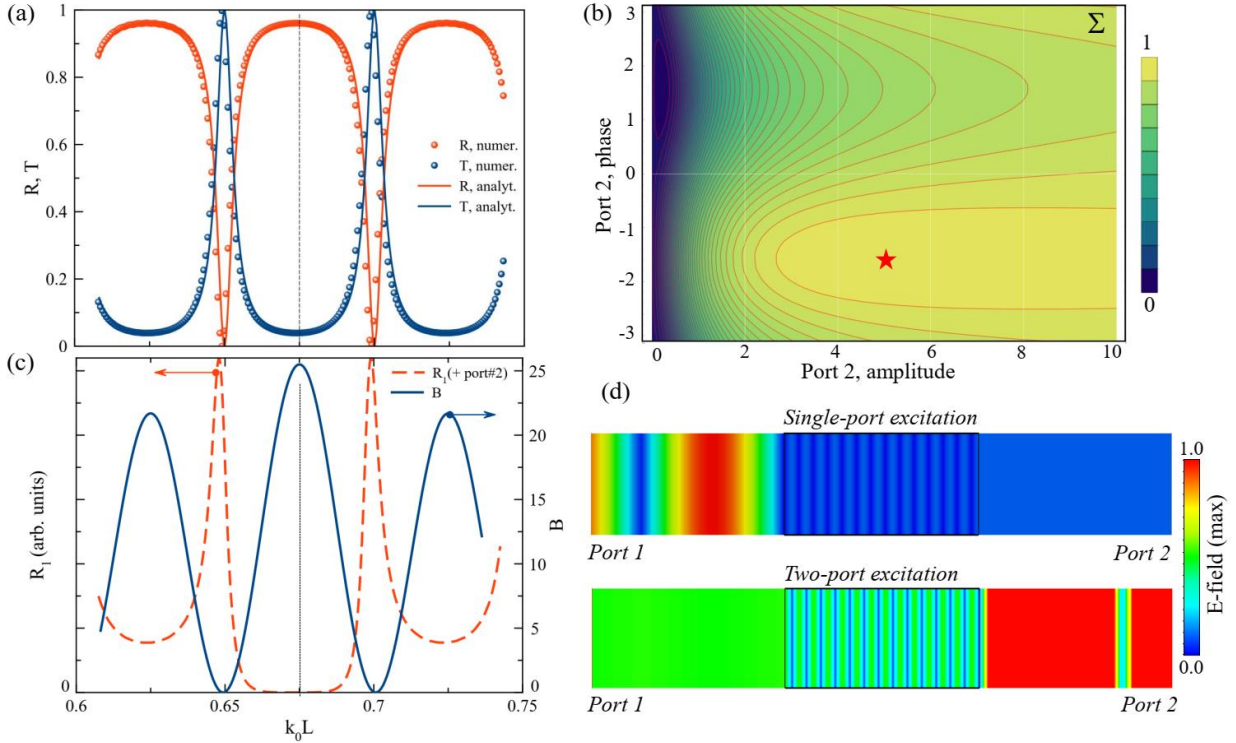

**Figure 3**. Coherently assisted power transfer enhancement through high-Q Fabry-Perot cavity made of a dielectric slab with relative dielectric permittivity $\varepsilon = 100$ situated in the air. The thickness $L$ is 405 nm. (a) One-port reflection (R) and transmission (T) coefficients of this slab calculated analytically (solid curves) and numerically (dots). (b) Normalized transferred power ($\Sigma$) calculated with Eq.(2) vs. the amplitude and phase of the auxiliary wave at port #2; the red star corresponds to the optimized regime. (c) Power transfer enhancement (blue curve) and the reflection from the port #1 (red dushed curve) in the optimized regime as a function of normalized frequency, $k_0 L$. (d) Normalized E-field distribution (max value over the period) for one-port excitation (upper) and two-port excitation (lower). The wave from port#1 in the optimized regime of two-port excitation propagates through the cavity without reflection.

Next, we calculate the normalized transferred power ($\Sigma$) as a function of the amplitude and phase of the auxiliary wave at port #2 by using Eq.(2), Figure 3(b). The red star corresponds to the optimized regime ($a = 5$, $\varphi = -1.5$), where the transferred energy is 1. Thus, in this regime, all energy from both ports ($|s_1^+|^2 + |s_2^+|^2$) went to port #2. The result of the power transfer enhancement (B) in this regime along with the corresponding reflection coefficient to port #1 (red dushed curve) as a function of normalized frequency, $k_0 L$ are presented in Figure 3(c). The result shows that the power transfer in this specific case is enhanced by 25 times. We see that in this optimized regime, the reflection coefficient to port #1 is vanishing and hence the wave propagating from the port #1 experience no back reflection; all energy goes towards the port #2. This conclusion is supported by the E-field distribution (max value over the period), Figure 3(d). In the case of two-port excitation (lower panel), we observe no back reflection, and all energy goes through the system towards port #2 despite its very small transmittance. Otherwise, for one-port excitation, we observe strong reflection with the formation of a standing wave (top panel).

It is worth noting that although this regime of *coherent perfect transmission* requires higher amplitudes of the auxiliary wave, the significant enhancement in the power transfer efficiency can be achieved at much weaker amplitudes. For example, the power transfer enhancement (B) of 10 can be achieved for $a \approx 0.5$. Also, it should be stressed that all energy of the auxiliary wave returns back to port #2 in the perfect transmission mode ($\Sigma = 1$) and hence there is no energy loss. Thus, this simple example shows that the energy can be transferred through arbitrary poorly transparent barrier via auxiliary coherent excitation with a certain amplitude and phase, which can be found calculating or measuring the transmission and reflection $r_{22}$, according to Eq. (2).

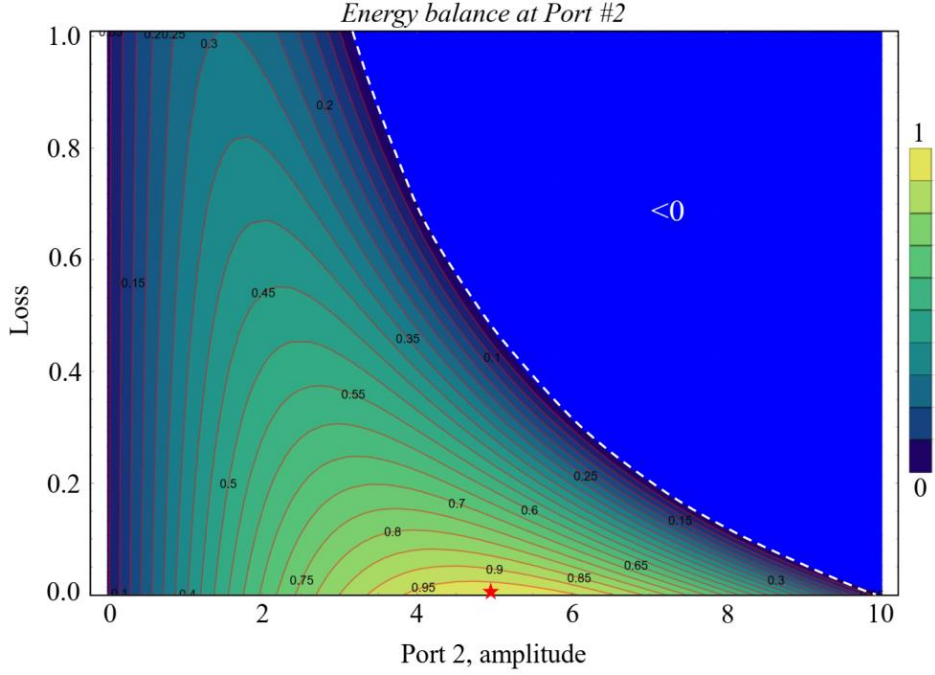

**Figure 4**. Energy balance at port #2 for Fabry-Perot cavity made of a dielectric slab with relative dielectric permittivity $\varepsilon = 100 + i \cdot \text{Loss}$ situated in the air as a function of loss factor and the amplitude of the auxiliary wave (the phase is -1.55). The thickness $L$ is 405 nm. The blue area corresponds to the negative energy balance. The star shows the optimized lossless case of Figure 3.

To discuss the lossy system case, we in Figure 4 we plot the energy balance at port #2, $|s_1^+ t + s_2^+ r_{22}|^2 - |s_2^+|^2$. If this value is positive, then the net energy flow is larger than the energy of the auxiliary wave ($|s_2^+|^2$) and goes to out from the system. This simply means that in the case of positive balance we get more energy at port #2 than spend for exciting it (the difference stems from port #1). Otherwise, when the energy gets dissipated and scattered to port #2, the energy balance gets smaller and eventually becomes negative. Figure 4 shows that at $a = 5$ (phase is -1.55) and $\text{Loss} = 0$, the energy balance equals 1, which corresponds to the full transmission of energy from port #1 to port #2. The enhancement of loss in the system leads to redusing of the balance, however, event at relatively high loss values, the balance stays positive giving rise to coherently enhanced energy transmission.

Finally, we consider a more complicated case of the wave transfer through a barely transparent metasurface made of the perfect electric conductor (PEC) with periodically perforated rectangular holes, Figure 5(a, inset). The period of this array is assumed to be $D = \lambda_0 / 6$, the size of the hole is $s = \lambda_0 / 7.5$, and the thickness is of $t = \lambda_0 / 20$. In our simulations, the actual parameters were $D = 100$ nm, $s = 80$ nm, and $t = 30$ nm and the central wavelength in vacuum $\lambda_0 = 600$ nm.

The calculation results of reflection (R) and transmission (T) coefficients for single-port excitation for this metasurface are presented in Figure 5(a). We see that over the frequency range of $k_0 D = 0.15 \div 0.18$ the surface has the large reflection coefficient of ~0.999 (red curve) and consequently vanishing transmission coefficient (blue curve).

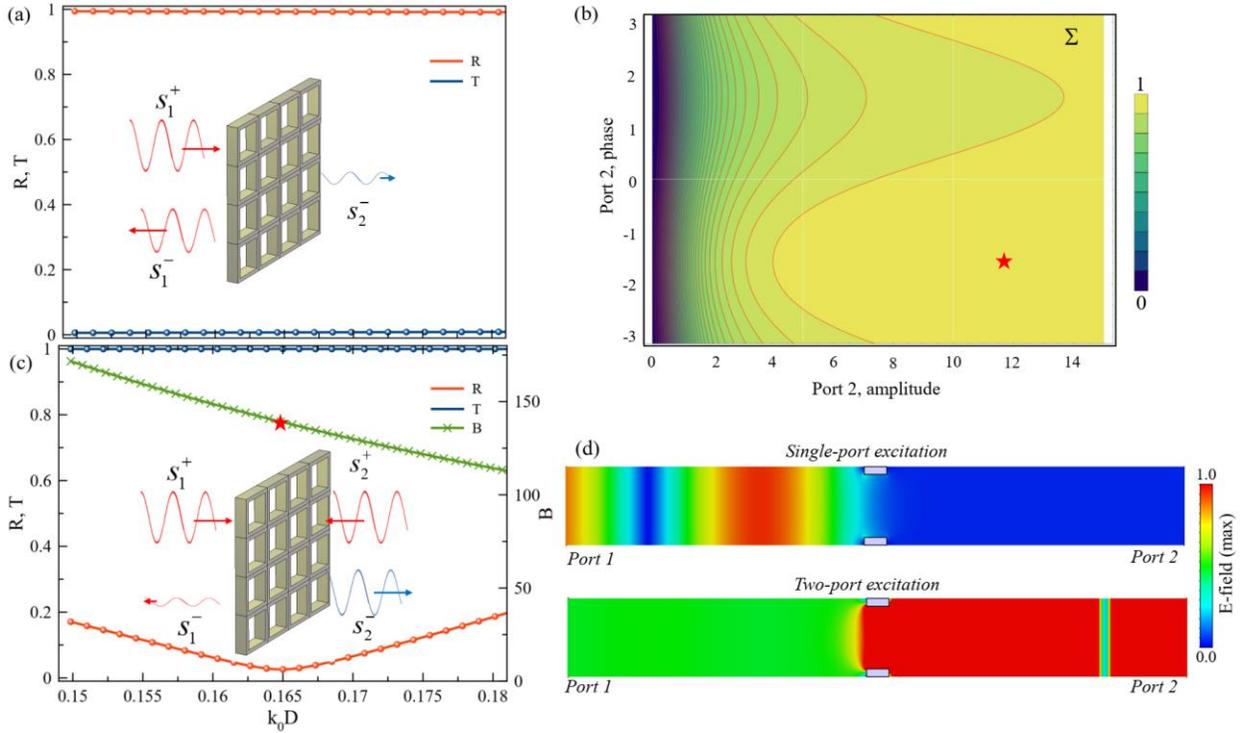

**Figure 5**. Coherently assisted power transfer enhancement through a barely transparent metasurface made of the perfect electric conductor (PEC) with periodically perforated rectangular holes. The period of this array is $D = \lambda_0 / 6$, the size of the hole is $s = \lambda_0 / 7.5$, and the thickness of $t = \lambda_0 / 20$. (a) Transmission and reflection for one-port excitation. (b) Normalized transferred power ($\Sigma$) calculated with Eq.(2) vs. amplitude and phase of the port #2; the red star corresponds to the optimized regime. (c) Coherently assisted power transfer enhancement (green), reflection (red) and transmission (blue) in the optimized regime as a function of normalized frequency, $k_0 D$. (d) E-field distribution (max value over the period) for one-port excitation (upper) and for two-

port excitation (lower). The actual parameters were $D = 100$ nm, $s = 80$ nm, and $t = 30$ nm and the central wavelength in vacuum $\lambda_0 = 600$ nm.

The calculation results of normalized transferred power ($\Sigma$) through this structure is presented in Figure 5(b) as a function of the amplitude and phase of the auxiliary wave at $k_0 D = 0.165$. The red star shows the optimized regime ($\Sigma = 1$) where the enhancement reaches 130 [Figure 5(c)] and corresponds to full transmission with vanishing reflection, Figure 5(c, red curve). If we would not know that there is an auxiliary wave at the opposite side of the structure, we would decide that the structure is transparent in this frequency range. However, as we remove the auxiliary wave, the energy from port #1 is almost entirely back reflected. Note that significant enhancement factors can be achieved at much smaller amplitudes of the auxiliary wave, e.g., for $a = 4$, the enhancement of ~110 is achievable, which is sufficient for practical applications.

## Conclusions

We have proposed a novel approach to efficient wide-band wireless power transfer through barely transparent barriers. The approach is based on coherent excitation of a barrier by a specially designed auxiliary wave with a certain amplitude and phase from the side where the energy should be delivered to. This auxiliary wave not only helps the initial signal travel through the barrier but also gets fully reflected back ensuring 100% transfer efficiency in a lossless case. We have analyzed the issue of energy transfer through barriers analytically in terms of S-matrix and derive the general equation, which allows treating the barrier of any type in the same way. We have demonstrated the applicability of the proposed approach for two different barriers: a Fabry-Perot high-index dielectric resonator and more complicated nonresonant structure in the form of periodically arranged subwavelength holes (metasurface). We have also demonstrated the usefulness of the proposed coherent approach for lossy systems. In contrast to the traditional solution, this approach is general and applicable to lossless and lossy barriers in any spectral range.